\documentclass[12pt]{article}
\usepackage{graphicx}
\usepackage{amssymb}

\newcommand\pubnumber{}
\newcommand\pubdate{\today}

\def\iitm{Department of Physics, Indian Institute of Technology, Madras\\
Chennai, 600036, India}
\def\support{\footnote{Work supported by the UK-India Education and Research Initiative}}

\def\Title#1{\begin{center} {\Large #1 } \end{center}}
\def\Author#1{\begin{center}{ \sc #1} \end{center}}
\def\Address#1{\begin{center}{ \it #1} \end{center}}

\newcommand\pubblock{\rightline{\begin{tabular}{l} \pubnumber\\
         \pubdate  \end{tabular}}}
\newenvironment{Abstract}{\begin{quotation}  }{\end{quotation}}
\newenvironment{Presented}{\begin{quotation} \begin{center} 
             PRESENTED AT\end{center}\bigskip 
      \begin{center}\begin{large}}{\end{large}\end{center} \end{quotation}}

\def\beq{\begin{equation}}
\def\eeq#1{\label{#1}\end{equation}}
\def\eeqn{\end{equation}}

\def\beqa{\begin{eqnarray}}
\def\eeqa#1{\label{#1}\end{eqnarray}}
\def\eeqan{\end{eqnarray}}

\let\bar=\overbar

\def\Dslash{\not{\hbox{\kern-4pt $D$}}}
\def\dslash{\not{\hbox{\kern-2pt $\del$}}}

\def\msb{{\bar{\ssstyle M \kern -1pt S}}}

\begin{document}
\begin{titlepage}
\pubblock

\vfill
\Title{$CP$ content of $D^{0}\to h^{+}h^{-}\pi^{0}$}
\vfill
\Author{Jim Libby\support}
\Address{\iitm}
\vfill
\begin{Abstract}
Quantum-correlated $\psi(3770) \to D\bar{D}$ decays collected by the CLEO-c experiment are used to perform measurements of $F_+$, the fractional $CP$-even content of the self-conjugate decays $D \to \pi^+\pi^-\pi^0$ and $D \to K^+K^-\pi^0$.  Values of
$0.973 \pm 0.017$  and  $0.732 \pm 0.055$ are obtained for $\pi^+\pi^-\pi^0$ and $ K^+K^-\pi^0$, respectively.  The high $CP$-even content of  $D \to \pi^+\pi^-\pi^0$, in particular, makes this a promising mode for improving the precision on $\gamma$ and for measurements of $CP$ violation in $D$ decay.
\end{Abstract}
\vfill
\begin{Presented}
The 7th International Workshop on Charm Physics (CHARM 2015)\\
Detroit, MI, 18-22 May, 2015
\end{Presented}
\vfill
\end{titlepage}
\def\thefootnote{\fnsymbol{footnote}}
\setcounter{footnote}{0}
%
\section{Introduction}
\label{sec:intro}

A better determination of the unitarity triangle angle $\gamma= \arg({-V_{\rm ud}V^*_{\rm ub}/  V_{\rm cd}V_{\rm cb}^*})$ is required for testing the $CP$ violation mechanism within the Standard Model.  Sensitivity to $\gamma$
can be obtained by studying $CP$-violating observables in $B^\mp \to D K^\mp$ decays, where $D$ indicates a neutral charm meson reconstructed in a final state common to both $D^0$ and $\bar{D}^0$ mesons, including $CP$-eigenstates~\cite{GLW}. The current world average precision on $\gamma$ is much worse than that of the other angles of the unitarity triangle~\cite{PDG}. Therefore, including additional $D$-meson final states is desirable to reduce the statistical uncertainty on $\gamma$ at current and future facilities. 

In the case that the $D$ does not decay to a pure $CP$ eigenstate, information is required on the strong decay dynamics in order  to relate the $CP$-violating observables to $\gamma$. This information can be obtained from studies of quantum-correlated $D\bar{D}$  mesons produced in $e^{+}e^{-}$ collisions at an energy corresponding to the mass of the $\psi(3770)$~\cite{GGR,GGSZ,ANTONBONDAR,ATWOODSONI}. The decay $D \to \pi^+\pi^-\pi^0$ is a promising candidate to be added to the modes used in the $\gamma$ measurement.   
Its Dalitz plot  has been studied by the CLEO and BaBar collaboration using flavour-tagged $D^0$ decays and exhibits a strikingly symmetric distribution that suggests the decay may be dominated by a single $CP$ eigenstate~\cite{CLEOPIPIPI0,BaBarPIPIPI0}.  An isospin analysis \cite{BRIAN} of the amplitude model for $D^{0}\to \pi^+\pi^-\pi^0$ presented in Ref.~\cite{BaBarPIPIPI0} concludes that the final state is almost exclusively $I=0$. Therefore, given that the parity and $G$-parity of the three-pion final state is odd and $G=(-1)^{I}C$, the final state is expected to be $C=-1$ and $CP=+1$.  As its branching ratio of $1.43 \pm 0.06\%$~\cite{PDG} is significantly larger than those of the pure two-body $CP$-even modes, it has the potential to contribute strongly in any analysis making use of such decays. The channel $D\to K^+ K^- \pi^0$ is a similar, but less abundant, self-conjugate mode that has also attracted interest  \cite{CLEOKKPI0,BaBarKKPI0}.  These proceedings present the first analysis of these decays using quantum-correlated $D\bar{D}$ decays,  and measurements of their $CP$ content, making use of the CLEO-c $\psi(3770)$ data set.  These measurements allow the inclusive decays to be included in future $B^\mp \to DK^\mp$ analyses in a straightforward and model-independent manner, thus allowing for an improved determination of the angle $\gamma$. Further it has been shown that improved determinations of CP violating parameters in the charm sector can be made with these measurements \cite{THOMAS}.

These proceedings are based on Refs. \cite{NAYAK} and \cite{4PI} and  are structured as follows. Section~\ref{sec:cpcontent} describes how quantum-correlated $D$ decays are used to determine the $CP$ content. The results  are presented in Sect.~\ref{sec:results}. In Sect.~\ref{sec:implications}  the implications for the measurement of the unitarity triangle angle $\gamma$ are discussed. Section~\ref{sec:conc} gives the conclusions. 
    	
\section{Measuring the $CP$ content}\label{sec:cpcontent}

Consider a  $\psi(3770) \to D\bar{D}$ analysis in which the signal decay mode is $D \to h^+h^-\pi^0$.
Let $M^+$ designate the number of ``double-tagged'' candidates, after background subtraction, where one $D$ meson is reconstructed in the signal mode of interest, and the other is reconstructed in a $CP$-odd eigenstate.  The quantum-numbers of the $\psi(3770)$ resonance then require that the signal mode is in a $CP$-even state, hence the $+$ superscript.  The observable $M^-$ is defined in an analogous manner.   Let $S^+$ ($S^-$) designate the number of  ``single-tagged'' $CP$-odd ($CP$-even) candidates in the data sample, where a $D$ meson is reconstructed decaying to a $CP$ eigenstate, with no requirement on the final state of the other $D$ meson in the event.  The small  effects  of $D^0\bar{D}^0$ mixing are eliminated from the measurement \cite{NAYAK}.

On the assumption that for double-tagged candidates the reconstruction efficiencies of each $D$ meson are independent,  then the quantity $N^{+} \equiv M^+ / S^+$ has no dependence on the branching fractions or reconstruction efficiencies of the $CP$-eigenstate modes, and can be directly compared with the analogous quantity $N^-$ to gain insight into the $CP$ content of the signal mode.  The $CP$ fraction is defined
\begin{equation}
F_+ \equiv \frac{N^+}{N^+ \, + \, N^-} \; ,
\label{eq:Fplus}
\end{equation}
and is $1$ ($0$) for a signal mode that is fully $CP$-even ($CP$-odd).   The notation $F_+ (\pi^+\pi^-\pi^0)$ and  $F_+ (K^+K^-\pi^0)$ is used in the discussion when it is necessary to distinguish between the two final states. In addition, tagging the final state with 
$K^{0}\pi^{+}\pi^{-}$ and measuring the yield in bins of the $K^{0}\pi^{+}\pi^{-}$ Dalitz of plot for which the strong phase parameters of the decay are known \cite{CLEOKSPIPI} yields further sensitivity to $F_{+}$ \cite{4PI}.

Amplitude models of $D^0 \to \pi^+\pi^-\pi^0$ and $D^0 \to K^+K^-\pi^0$ are available from studies of flavour-tagged $D^0$ decays performed by the BaBar collaboration~\cite{BaBarPIPIPI0,BaBarKKPI0}.  These models can be used to calculate predictions for the $CP$ content for each decay.  Values of  $F_+ ({\pi^+\pi^-\pi^0}) = 0.92$ and   $F_+ ({K^+K^-\pi^0}) = 0.64$ are obtained.


\section{Results}\label{sec:results}
The data set analysed consists of $e^+e^-$ collisions produced by the Cornell Electron Storage Ring (CESR) at $\sqrt{s}=3.77$~GeV and collected with the CLEO-c detector. The integrated luminosity of the data set is 818~$\rm pb^{-1}$.  The CLEO-c detector is described in detail elsewhere~\cite{CLEOC}.  Simulated Monte Carlo (MC) samples of signal decays are used to estimate selection efficiencies. Possible background contributions are determined from a generic MC sample corresponding to
approximately ten times the integrated luminosity of the data set.  The EVTGEN generator~\cite{EVTGEN} is used to simulate the decays.  The detector response is modelled using the  GEANT software package~\cite{GEANT}.

Table~\ref{tab:finalstates} lists the reconstructed $D^{0}$ and
$\bar{D}^{0}$ final states. The unstable final state particles are reconstructed in the following decay modes: $\pi^{0}\to\gamma\gamma$, $K^{0}_{\rm S}\to\pi^{+}\pi^{-}$,
$\omega\to\pi^+\pi^-\pi^0$, 
$\eta\to\gamma\gamma$, $\eta\to\pi^+\pi^-\pi^0$ and
$\eta^{\prime}\to\eta(\gamma\gamma)\pi^{+}\pi^{-}$. The $\pi^{0}$, $K^{0}_{\rm S}$, $\omega$, $\eta$ and $\eta^{\prime}$  
reconstruction procedure is identical to that used in Ref. \cite{WINGS}. 

\begin{table}[th]
\begin{center}
\caption{$D$ final states reconstructed in this analysis.} \vspace*{0.1cm}
\label{tab:finalstates}
\begin{tabular}{cc}\hline\hline
Type & Final states \\ \hline
Signal & $\pi^+\pi^-\pi^0$, $K^+K^-\pi^0$ \\
$CP$-even & $K^{+}K^{-}$, $\pi^{+}\pi^{-}$, $K^{0}_{\rm S}\pi^{0}\pi^{0}$, $K^{0}_{\rm L}\pi^{0}$, $K^{0}_{\rm L}\omega$ \\
$CP$-odd  & $K^{0}_{\rm S}\pi^{0}$, $K^{0}_{\rm S}\omega$, $K^{0}_{\rm S}\eta$, $K^{0}_{\rm S}\eta^{\prime}$ \\
Mixed CP & $K^{0}_{\rm S}\pi^{+}\pi^{-}$, $K^{0}_{L}\pi^{+}\pi^{-}$ \\
\hline   
\hline   
\end{tabular}
\end{center}
\end{table}

Final states that do not contain a $K^{0}_{\rm L}$ are fully reconstructed via two kinematic variables: the 
beam-constrained candidate mass, $M_{bc}\equiv\sqrt{s/4c^{4}-\mathbf{p}_{D}^{2}/c^{2}}$, where 
$\mathbf{p}_{D}$ is the $D$-candidate momentum, and $\Delta E\equiv E_{D}-\sqrt{s}/2$, where $E_{D}$ is the 
$D$-candidate energy. The $M_{bc}$ and $\Delta E$ distributions of correctly reconstructed $D$-meson candidates will peak at the nominal $D^{0}$ mass and zero, respectively. Neither $\Delta E$ nor $M_{bc}$ distributions exhibit any peaking structure for combinatoric background. The double-tagged yield is determined from counting events in signal and
sideband regions of $M_{bc}$ after placing requirements on $\Delta E$. The yield determination procedure is identical to that presented in Refs.~\cite{TQCA1,WINGS}. 

\begin{figure}[th]
\begin{center}
\begin{tabular}{cc}
\includegraphics[width=0.45\columnwidth]{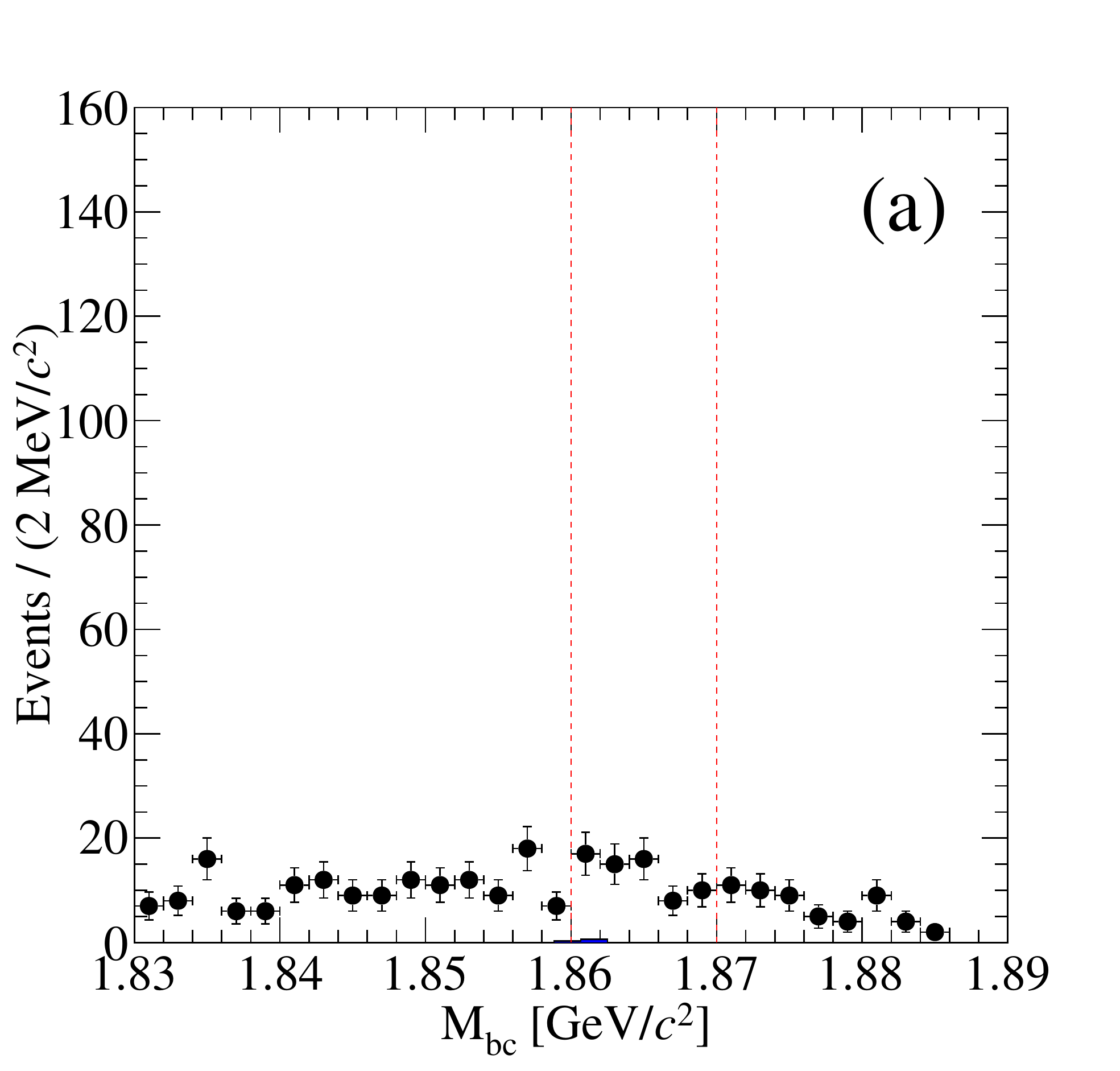} &
\includegraphics[width=0.45\columnwidth]{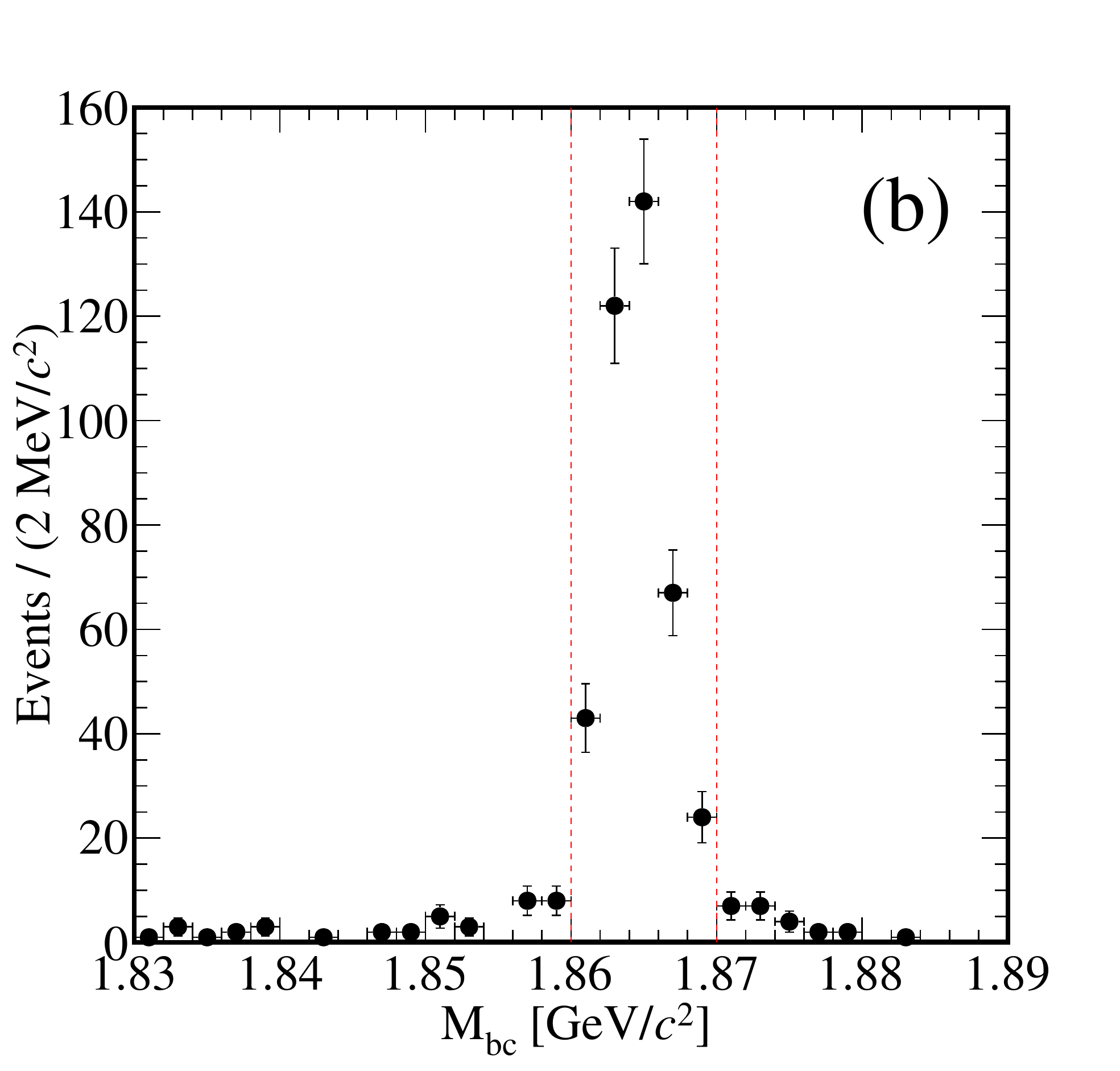}
\end{tabular}
\caption{$M_{bc}$ distributions for $D \to\pi^+\pi^-\pi^0$ candidates tagged by $CP$-even (a) and  $CP$-odd (b) eigenstates. Tags involving a $K^0_{\rm L}$ are not included.
The vertical dotted lines indicate the applied signal window.}  \label{fig:hhpi0_signal_mbc}
\end{center}
\end{figure}

The selection procedures are almost identical to those presented in Refs.~\cite{TQCA1,WINGS}; additional details of the selection can be found in Ref.~\cite{NAYAK}.  Figure~\ref{fig:hhpi0_signal_mbc} shows the $M_{bc}$ distributions for $CP$-tagged signal candidates, summed over all $CP$-even and $CP$-odd tags, respectively, where the $CP$-tag final state does not contain a $K^{0}_{\rm L}$ meson. No significant signal is seen in any of the modes tagged by a $CP$-even eigenstate, whereas significant signals are seen in most modes tagged by $CP$-odd eigenstates.

\begin{figure} 
\begin{center}
\begin{tabular}{c}
\includegraphics[width=0.45\columnwidth]{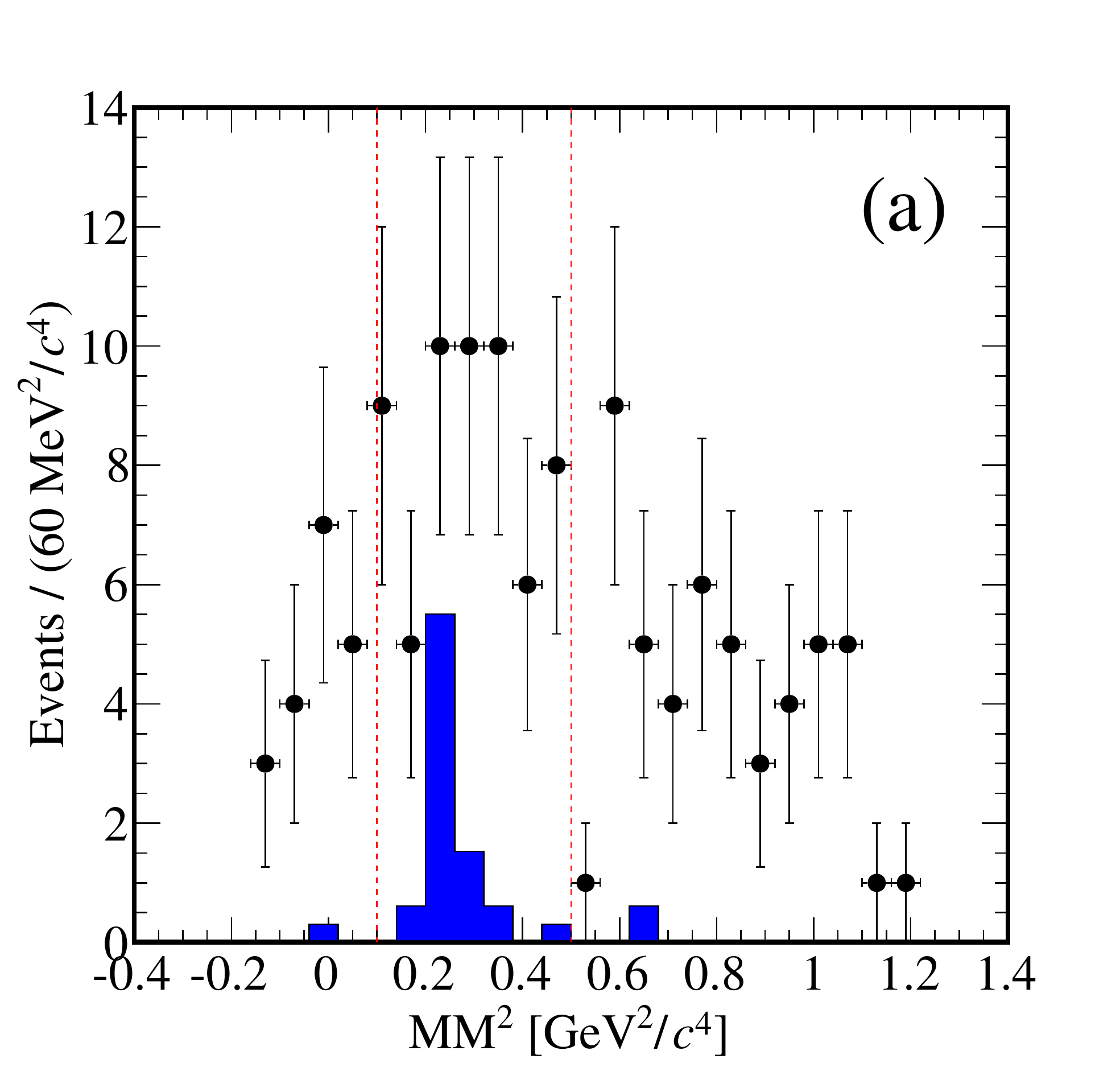} 
\end{tabular}
\caption{$M_{\mathrm{miss}}^2$ distributions for $D \to\pi^{+}\pi^{-}\pi^{0}$. The shaded histogram indicates the peaking background. The vertical dotted lines indicate the applied signal window.} \label{fig:hhpi0_signal_MM2}
\end{center}
\end{figure}

Many $K^{0}_{\rm L}$ mesons produced do not deposit any reconstructible signal in the detector. However, double-tag candidates can be fully reconstructed using a missing-mass squared $(M_{\mathrm{miss}}^2)$ technique \cite{K0LPRL} for tags containing a single $K^{0}_{\rm L}$ meson. Yields are extracted from the signal and sideband regions of the
$M_{\mathrm{miss}}^2$ distribution. Figure~\ref{fig:hhpi0_signal_MM2}  shows the $M_{\mathrm{miss}}^2$ distributions for candidates tagged with either a $K^0_{\rm L} \pi^0$ or $K^0_{\rm L} \omega$ tag.

It is also necessary to know the single-tag yield for the $CP$-eigenstates to normalise the double-tagged yields appropriately to obtain a value of $F_{+}$. The details of this selection can be found in Ref.~\cite{NAYAK}

The yields of double-tagged and single $CP$-tag candidates are used to determine the quantities $N^+$ and $N^-$, and from these the $CP$ fraction $F_+$.  The values for $N^+$ and $N^-$ are calculated from the ensemble of $CP$-odd and $CP$-even tags, respectively, accounting for statistical and systematic uncertainties, and allowing for the correlations that exist between certain systematic components.

\begin{figure}[th]
\begin{center}
\begin{tabular}{cc}
\includegraphics[width=0.45\columnwidth]{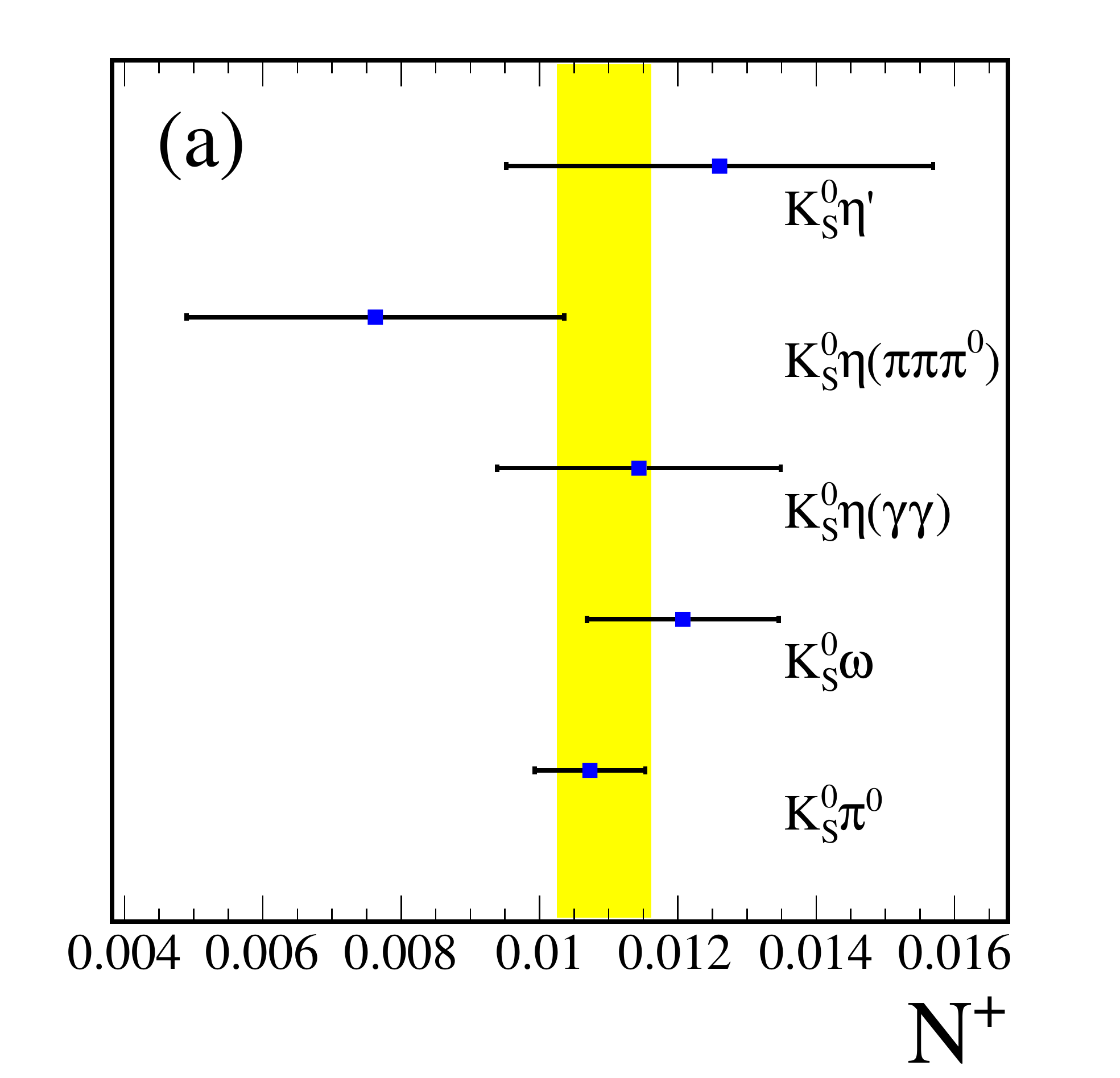} &
\includegraphics[width=0.45\columnwidth]{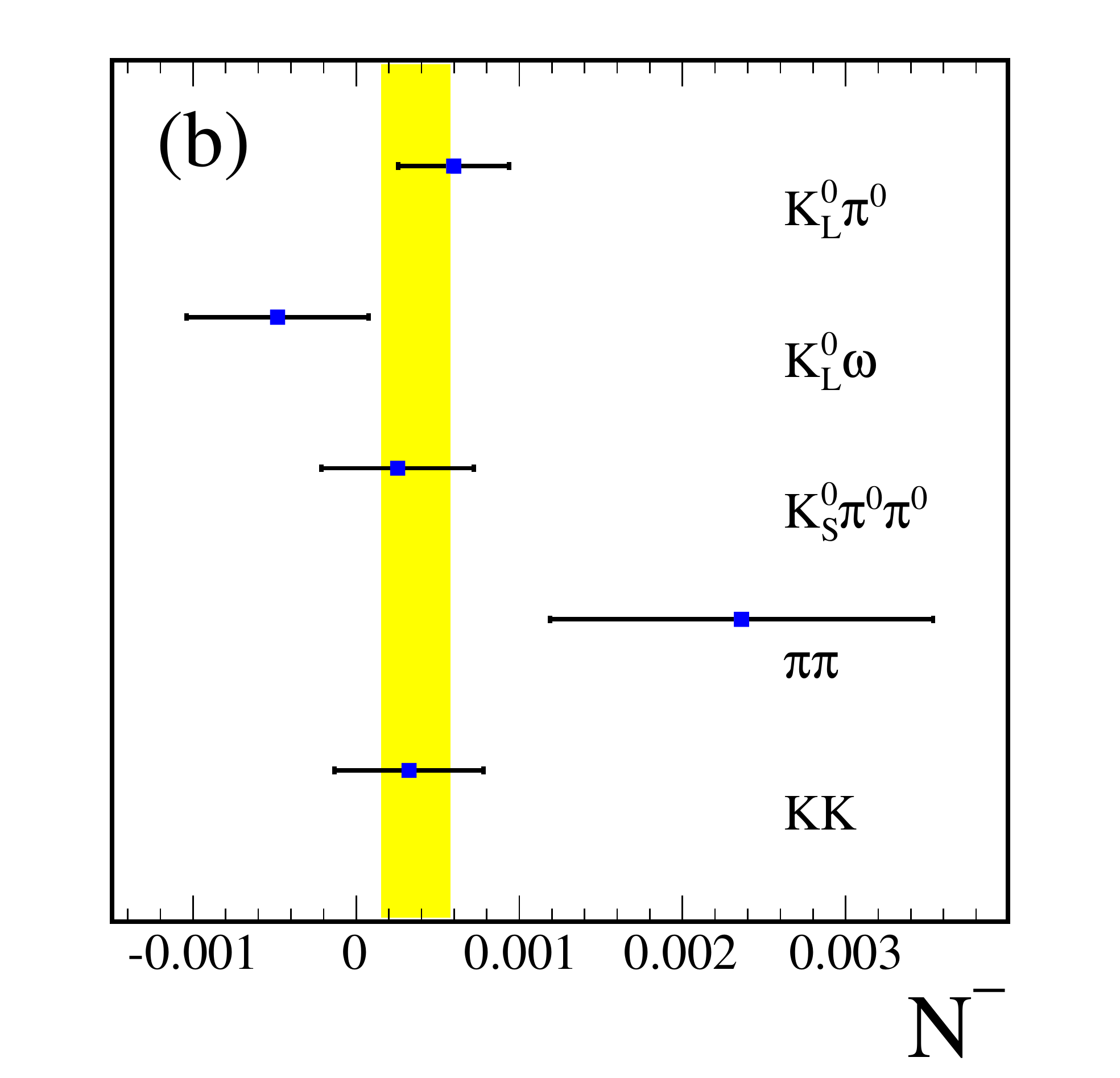} \\
\end{tabular}
\caption{$D \to \pi^+\pi^-\pi^0$ results for $N^+$ (a) and $N^-$ (b).  
In each plot the vertical yellow band indicates the value obtained from the combination of all tags.  }  \label{fig:results}
\end{center}
\end{figure}

The measured values for $N^+$ and $N^-$ for the two signal modes are displayed in Fig.~\ref{fig:results}.  It can be seen that there is consistency between the individual $CP$ tags for each measurement.  From these results it is determined that
$F_+ ({\pi^+\pi^-\pi^0}) = 0.968 \pm 0.017 \pm 0.006$  and  $F_+({K^+K^-\pi^0}) = 0.731 \pm 0.058 \pm 0.021$, where the first uncertainty is statistical and the second is systematic. In addition, the binned yields from the $K^{0}_{L,S}\pi^{+}\pi^{-}$ tagged are used to determine values of $F_+$ which are found to be consistent with those from the $CP$ tags alone \cite{4PI}. The combined values are $0.973 \pm 0.017$  and  $0.732 \pm 0.055$ for $\pi^+\pi^-\pi^0$ and $ K^+K^-\pi^0$, respectively \cite{4PI}. These values are slightly higher than, but compatible with, the model predictions  reported in Sect.~\ref{sec:cpcontent}.
%
%

\section{Implications for the measurement of  $\boldmath \gamma$} 
\label{sec:implications}

Sensitivity to the unitarity triangle angle $\gamma$ is obtained by measuring the relative rates of $B^\mp \to D({h^+h^-\pi^0}) K^\mp$ decays and related observables. These partial widths and those involving flavour-specific $D$ meson decays can be used to construct the  partial-widths ratio $R_{F_+}$ and $CP$-asymmetry $A_{F_+}$:
\begin{eqnarray}
R_{F_+}  & \equiv & 
\frac{ \Gamma(B^-\to D_{F_+} K^- ) \,+\,  \Gamma(B^+ \to D_{F_+} K^+ ) }
{ \Gamma(B^-\to D^0 K^- ) \,+ \,  \Gamma(B^+ \to \bar{D}^0 K^+ ) }, \\
A_{F_+} & \equiv & 
\frac{ \Gamma(B^-\to D_{F_+} K^- ) \,-\,  \Gamma(B^+ \to D_{F_+}K^+) }
{ \Gamma(B^-\to D_{F_+} K^- ) \,+ \,  \Gamma(B^+ \to D_{F_+ } K^+)} , 
\end{eqnarray}
where $D_{F_+}$ indicates a $D$ meson of $CP$-even content $F_+$, established through its decay into the final state $h^+h^-\pi^0$.  These observables are directly analogous to the usual so-called GLW~\cite{GLW} observables  $R_{{\rm CP }\pm}$ and $A_{{\rm CP} \pm}$, where the $D$ meson is reconstructed in a pure $CP$ eigenstate.

   Then  $R_{F_+}$ and $A_{F_+}$ are found \cite{NAYAK} to have the following dependence on the underlying physics parameters:
\begin{eqnarray}
R_{F_+}  & = & 1 \, + \, r_B^2 + (2F_+ -1) \cdot 2r_B\cos\delta_B \cos \gamma,  \\
A_{F_+} &=& (2F_+ -1 ) \cdot 2r_B \sin\delta_B \sin\gamma / R_{{F_+}},
\end{eqnarray}
which reduces to the equivalent expressions for  $R_{{\rm CP }\pm}$ and $A_{{\rm CP} \pm}$ in the case $F_+$ is $1$ or $0$.  Therefore inclusive final states such as $h^+h^-\pi^0$ may be cleanly interpreted in terms of $\gamma$ and the other parameters of interest, provided that $F_+$ is known.  At leading order the only difference that the $CP$ asymmetry $A_{F_+}$ has with respect to the pure $CP$-eigenstate case is a dilution factor of $(2F_+ -1)$.

\section{Conclusion}\label{sec:conc}
Data corresponding to an integrated luminosity of 818~$\rm pb^{-1}$ collected by the CLEO-c experiment in $e^+e^-$ collisions at the $\psi(3770)$ resonance have been analysed for the decays $D \to \pi^+\pi^-\pi^0$ and $D \to K^+K^-\pi^0$.  Measurements of  $F_+$, the fractional $CP$-even content of each decay have been performed showing that $D^{0}\to\pi^+\pi^-\pi^0$ is nearly a pure $CP$-even eigenstate.
It has been demonstrated that such self-conjugate inclusive channels can be cleanly included in measurements of the unitarity-triangle angle $\gamma$, using $B^\mp \to D K^\mp$ decays.  The high value of $F_+$  obtained for  $D \to \pi^+\pi^-\pi^0$ makes this channel, in particular, a valuable addition to the suite of $D$-decay modes used in the measurement of $\gamma$ at LHCb and Belle-II.  Improved precision on the $F_+$ parameters can be obtained using the larger $\psi(3770)$ data set available at BESIII, and similar measurements can also be performed for other self-conjugate final states.


\section*{Acknowledgments}

This analysis was performed using CLEO-c data, and as a member of the former CLEO collaboration I thank it for this privilege. I am grateful for support from the UK-India  Education and Research Initiative.





\end{document}